\documentstyle[sprocl,epsf]{article}

\def\Journal#1#2#3#4{{#1} {\bf #2}, #3 (#4)}

\def\NPB{{\em Nucl. Phys.} B}
\def\PLB{{\em Phys. Lett.}  B}

\def\PRD{{\em Phys. Rev.} D}
\def\AP{{\em Ann. Phys.}}

\def\ra{\rightarrow}

\def\be{\begin{equation}}
\def\ee{\end{equation}}
\def\bea{\begin{eqnarray}}
\def\eea{\end{eqnarray}}

\def\beeq{\begin{equation}}
\def\eneq{\end{equation}}
\def\beqn{\begin{eqnarray}}
\def\eeqn{\end{eqnarray}}

\def\mybig{\displaystyle \strut }

\def\dd{\partial}
\def\la{\raise.16ex\hbox{$\langle$}\lower.16ex\hbox{}  }
\def\ra{\, \raise.16ex\hbox{$\rangle$}\lower.16ex\hbox{} }
\def\go{\rightarrow}

\def\onehalf{ \hbox{${1\over 2}$} }

\def\psibar{ \psi \kern-.65em\raise.6em\hbox{$-$} }
\def\chibar{ \chi \kern-.65em\raise.5em\hbox{$-$} }
\def\mbar{ m \kern-.75em\raise.4em\hbox{$-$}\hbox{} }
\def\Bbar{ B \kern-.73em\raise.6em\hbox{$-$}\hbox{} }

\def\ep{\epsilon}

\def\wil{ \Theta_{\rm W} }

\def\vphi{ {\varphi} }

\def\tot{{\rm tot}}

\def\eff{{\rm eff}}
\def\mass{{\rm mass}}

\def\LapN{{\triangle_N}}
\def\potN{{V_N}}
\def\myfrac#1#2{{\mybig #1 \over \mybig #2}}

\def\boxit#1{$\vcenter{\hrule\hbox{\vrule\kern3pt
     \vbox{\kern3pt\hbox{#1}\kern3pt}\kern3pt\vrule}\hrule}$}
\def\bigbox#1{$\vcenter{\hrule\hbox{\vrule\kern5pt
     \vbox{\kern5pt\hbox{#1}\kern5pt}\kern5pt\vrule}\hrule}$}
\def\hugebox#1{$\vcenter{\hrule\hbox{\vrule\kern8pt
     \vbox{\kern8pt\hbox{#1}\kern8pt}\kern8pt\vrule}\hrule}$}

\begin{document}

\rightline{\small UMN-TH-1430/96}
\rightline{\small AZPH-TH/96-15}
\vglue .5cm

\title{CONFINEMENT AND CHIRAL DYNAMICS IN THE MULTI-FLAVOR 
SCHWINGER MODEL\footnote{~To appear in the Proceedings of 
{\it Continuous Advances in QCD 96}, University of Minnesota, March
28-31, 1996.}}

\author{Y. HOSOTANI AND R. RODRIGUEZ}

\address{School of Physics and Astronomy, University of Minnesota\\
Minneapolis, MN 55455, USA}

\author{J.E. HETRICK}

\address{Department of Physics, University of Arizona,  Tucson,
AZ 85721, USA}

\author{S. ISO}

\address{Theory Division, KEK, Tsukuba,
Ibaraki 305, Japan}


\maketitle\abstracts{
Two-dimensional QED with $N$ flavor fermions is solved at zero and finite
temperature with arbitrary fermion masses to explore QCD physics such
as chiral condensate and string tension.  The problem is reduced to solving 
a Schr\"odinger equation for $N$ degrees of freedom with a specific
potential determined by the ground state of the Schr\"odinger problem
itself.}

\section{QCD$_4$ physics in QED$_2$}

Two-dimensional QED with massive fermions is not exactly solvable,
ie. it is not reduced to a free theory like the massless case. When
the fermion masses are large it becomes a highly interacting model.
Due to its many similarities with four-dimensional QCD in such respects
as instantons, chiral dynamics, and confinement, it has long been a
testing ground for intuitive physical principles and new ideas.

One of the purposes of this work is to clarify these aspects of QCD
physics by evaluating chiral condensates, the Polyakov loop, and
string tension as fermion masses ($m$), vacuum angle ($\theta$), and
temperature ($T$) vary.  We shall show that $N$ flavor QED is
effectively reduced to the quantum mechanics of $N$ degrees of freedom,
which can be solved numerically on workstations.\cite{HHI1,RH}

The Lagrangian is 
\beeq
{\cal L} = - \hbox{$1\over 4$} \, F_{\mu\nu} F^{\mu\nu} + 
\sum_{a=1}^N \psibar_a \Big\{ \gamma^\mu (i \dd_\mu - e A_\mu) -
  m_a  \Big\} \psi_a ~~.
\eneq
We examine the model defined on a circle $S^1$ with a circumference
$L$.  Upon imposing periodic and anti-periodic boundary conditions on
the bosonic and fermionic fields, respectively, the model is
mathematically equivalent to a theory defined on a line ($R^1$) at
finite temperature ($T$) by analytic continuation to imaginary time
$\tau$ $(=it$) and interchange of $\tau$ and $x$.  Various physical
quantities at $T\not=0$ on $R^1$ are obtained from the corresponding
ones at $T=0$ on $S^1$ by substituting $T^{-1}$ for $L$.

\section{Zero-mode Hamiltonian}

To begin the computation we bosonize the fermions. Each
two-component fermion ($\psi_a$) is expressed in terms of zero modes
($q_a^\pm , p_a^\pm$) and oscillatory modes ($\phi_a(x), \Pi_a(x)$).
The boundary conditions enforce that $p_a^\pm$'s take integer
eigenvalues.  $p_a^- - p_a^+$ and $p_a^- + p_a^+$ correspond to charge
and chiral charge, respectively.  In a vector-like theory without
background charges one can stay in a subspace defined by
$p_a^+=p_a^-$.

The relevant parts of the Hamiltonian are expressed in terms of
($q_a$=$q_a^++q_a^-, p_a$=$\onehalf[p_a^++p_a^-]$), ($\phi_a,\Pi_a$),
and ($\wil, P_W$) where $\wil$ is the Wilson line phase, the only
physical degree of freedom associated with gauge fields on a circle.
The Hamiltonian becomes
\beqn
&&H_\tot = H_0 + H_\phi + H_\mass\cr 
\noalign{\kern 8pt}
&&H_0 = {\pi \mu^2 L\over 2N} P_W^2 
+ {1\over 2\pi L} \sum_{a=1}^N (\Theta_W+2\pi p_a)^2 \cr
\noalign{\kern 5pt}
&&H_\phi = \int_0^L dx \, {1\over 2} 
\bigg\{  \sum_{a=1}^N \Big( {\Pi_a}^2 +\phi_a'^2 \Big)  
+ \mu^2 \,  \Big( {1\over \sqrt{N}}\sum_a \phi_a \Big)^2
 \bigg\}
\label{Hamiltonian1}
\eeqn
where $\mu^2=Ne^2/\pi$.  $H_\mass$ represents the contribution coming
from the fermion mass terms.  Expression (\ref{Hamiltonian1}) is
exact, from which it immediately follows that QED$_2$ with massless
fermions is exactly solvable. It contains 1 massive boson and $N-1$
massless bosons.

The fermion masses give nontrivial interactions among the zero modes
and $\phi$ modes, and the previously massless bosons now become
massive.  To find the true vacuum, we first determine the vacuum wave
function in the zero mode sector with given physical boson masses
$\mu_\alpha$'s ($\alpha$=1,$\cdots, N$). The boson masses, which
depend on the vacuum structure in the zero mode sector, are recomputed
with the vacuum wave function thus obtained. Since input and output values
for the boson masses must be the same, this gives a self-consistency
condition which we can solve for numerically in general, and
analytically in certain limits.

As a basis one may take eigenstates of ($P_W,p_a$) with eigenvalues
($p_W, n_a$) where the $n_a$'s are integers.
It is more convenient however to take a coherent
state basis with respect to $\{ n_a \}$. The vacuum wave function
is written as $\hat f(p_W; \vphi_1, \cdots, \vphi_{N-1}; \theta)$.  One
of the angular variables, $\theta$, specifies the so-called
$\theta$-vacuum; due to gauge invariance its value does not change.

When fermion masses are small, $\hat f = e^{-\pi \mu L p_W^2/2N}
f(\vphi;\theta)$ to good accuracy.  $f(\vphi;\theta)$ must satisfy
\beqn
&&\Big\{ -\LapN + \potN \Big\} ~ f(\vphi_1, \cdots, \vphi_{N-1})
= \ep ~ f(\vphi_1, \cdots, \vphi_{N-1}) \cr
\noalign{\kern 6pt}
&&\LapN = 
\sum_{a=1}^{N-1} {\dd^2\over \dd\vphi_a^2} 
-{2\over N-1} \sum_{a<b}^{N-1} {\dd^2\over \dd\vphi_a\dd\vphi_b} \cr
&&\potN = -
~  \sum_{a=1}^N   m_a A_a  \cos \vphi_a    \hskip .5cm   
\Big( \sum_{a=1}^N \vphi_a =\theta \Big) 
\label{Sch1}
\eeqn
where $A_a$ is determined by the boson masses $\mu_\alpha$. The
problem is now a Schr\"odinger equation. The salient feature is that
the potential has to be determined self-consistently such that its
ground state wave function (and hence the boson masses $\mu_\alpha$)
reproduces the same potential:
\beeq
 \potN(\vphi) \go f(\vphi) \go \mu_\alpha \go \potN(\vphi) ~. 
\eneq

\section{Chiral dynamics}
It is straightforward to determine the chiral condensate. In the large
or small volume limit (or equivalently in the low or high temperatue
limit) an analytic expression can be obtained. For $N\ge 3$ with
degenerate fermion masses ($m_a=m \ll \mu$)
\beeq
{\mybig 1\over\mybig  \mu} \la \psibar\psi \ra_\theta =
\cases{
- {\mybig 1\over\mybig  4\pi}
\Big(2 e^\gamma \cos {\mybig \bar\theta\over\mybig  N} \Big)^{{2N\over N+1}} \,
 \Big( {\mybig m\over \mybig \mu} \Big)^{{N-1\over N+1}}
   &for $T \ll m^{{N\over N+1}} \mu^{{1\over N+1}}$ \cr
- {\mybig 2N \over\mybig  \pi(N-1)} \, {\mybig m\over \mybig \mu} \, 
 e^{-2\pi T/N\mu}
   &for $T \gg \mu$~~.\cr} 
\eneq 
Here $\bar\theta$ is defined in the interval $-\pi\le \bar\theta \le +\pi$ by
$\bar\theta=\theta-2\pi[(\theta+\pi)/2\pi]$.

A few conclusions can be drawn here. First of all, at low $T$ the
chiral condensate is not analytic in $m$. This point was noticed in
the $N=2$ case at $T=0$ by Coleman twenty years ago.\cite{Coleman}
Secondly, at $T=0$, a cusp singularity appears at $\theta=\pi$.
Thirdly at high temperature perturbation theory in the fermion masses
is applicable since the condensate becomes analytic in $m$.

At moderate temperatures Eq.\ (\ref{Sch1}) must be solved numerically.
In fig.\ 1 we have displayed the $\theta$ dependence of chiral condensate
in the $N=3$ case with $m/\mu=0.01$ at various temperatures.  One can see 
the cusp singularity develop as $T$ approaches zero.

\begin{figure}[t]
\epsfxsize= 9.0cm    
\epsffile[100 240 450 500]{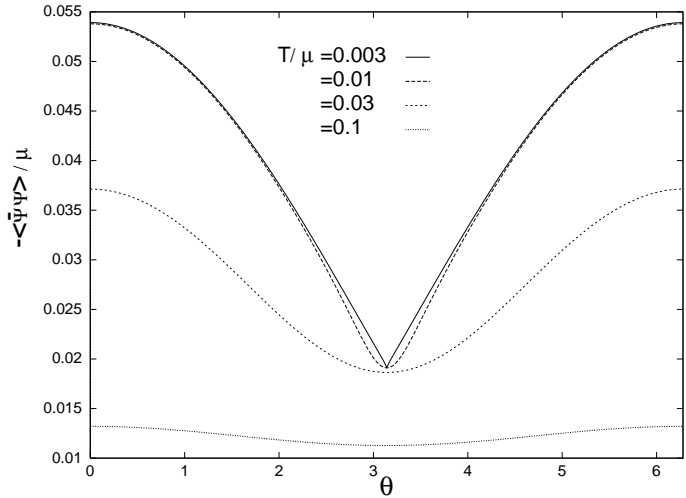}
\caption{$\theta$-dependence of the chiral condensate at various
temperatures in the $N=3$ model with $m/\mu=.01$.
At $T=0$, a cusp singularity appears at $\theta=\pi$.
\label{fig:theta}}
\end{figure}

At very low $T$ (large $L$) the potential term dominates over the kinetic
energy term in Eq.\ (\ref{Sch1}).  In other words, the ground state wave
function $f(\vphi)$ has a sharp peak around the location of the absolute
minimum of the potential $V_N(\vphi)$.  As $\theta$ varies, the location of
the  minimum also changes.  In the case of degenerate fermion masses, the 
minimum is located at $\vphi_a=\bar\theta/N$.  The location of the minimum
discontinuously shifts at $\theta=\pi$ (mod ~$2\pi$), which is the 
origin of the cusp singularity encountered in the $\theta$ dependence of
the chiral condensate at $T=0$. 

When the fermion masses are not degenerate, the coefficients $m_aA_a$
are all distinct.  In the three flavor ($N=3$) case the potential
takes the form
\beeq
V_3[\vphi] = -\Big\{ q_1 \cos \vphi_1 + q_2 \cos \vphi_2 
+ q_3\cos(\theta - \vphi_1 - \vphi_2) \Big\}
\label{potential1}
\eneq
where all $q_a$'s are different.  This potential has the same structure
as the effective chiral Lagrangian in QCD.  In the effective Lagrangian
written by Witten \cite{Witten} a potential term reads
\beeq
V^{\rm Witten}(U) = f_\pi^2 \Big\{ -{1\over 2} {\rm Tr}\, M(U+U^\dagger) 
 + {k\over 2N_c} (-i \ln \det U - \theta)^2 \Big\}
\label{witten}
\eneq
Here $U$ is the pseudoscalar field matrix, whereas
 $M= \, diag \, (m_u, m_d, m_s)$ is the quark mass matrix .
The second term represents the contribution from instantons.  The coefficient
$k$ is O(1) in the large $N_{\rm color}$ limit.

Phenomenologically $m_{\eta'}^2 \gg m_\pi^2, m_K^2, m_\eta^2$, which
implies that $k/N_c \gg m_a$, or that upon diagonalizing $U= \, {\rm
diag} \, (e^{i\phi_1}, e^{i\phi_2}, e^{i\phi_3})$, $\sum
\phi_a=\theta$.  Hence (\ref{witten}) reduces to (\ref{potential1}).
These equations are not exactly the same however, as $q_a$ in
(\ref{potential1}) is not simply $m_a$ but depends on $\{ m_a \}$
rather nontrivially.

In the past various conclusions were drawn based on (\ref{witten}),
which turn out to be perfectly correct in our context as well.
Physical quantities are periodic in $\theta$ with period $2\pi$.
With degenerate quark masses a cusp singularity appears at
$\theta=\pi$.  Furthermore sufficiently large asymmetry in quark
masses removes the singularity.\cite{Creutz}

Indeed, we have observed that at $T=0$ the location of the
minimum of the potential determines the vacuum wave function. 
When $q_1=q_2\ll q_3$, i.e.\ the strange quark is heavy but the up and
down quarks are degenerate ($m_u=m_d \ll m_s$), we find that
the discontinuous jump at $\theta=\pi$ in the location of the
minimum remains. However, if one adds small asymmetry in the up and down quark
masses ($m_u < m_d \ll m_s$), the minimum moves continuously to make a 
loop as $\theta$ changes from $-\pi$ to $+\pi$.  In other words
the cusp singularity disappears.

\section{Confinement}
One common method of examining confinement is to evaluate
the Polyakov loop.  In our method this is simply the 
expectation value of the Wilson line phase:
\beqn
P_q = \la e^{iq\int_0^\beta d\tau \, A_0(\tau,x)}\ra 
=e^{-\beta F_q}
\Leftrightarrow \la e^{i(q/e)\Theta_W(t)} \ra_{L=T^{-1}}&&\cr
\noalign{\kern 15pt}
=\cases{
0 &for $\myfrac{q}{e} \not=$ integer\cr
e^{-{\pi\mu n^2\over 4NT}} \int [d\vphi] \, 
  f(\vphi_a)^*f(\vphi_a+ {2\pi n\over N}) 
 &for $\myfrac{q}{e}=n$\cr} &&
\label{P-loop}
\eeqn

The result for $q=e$ is displayed in fig.\ 2.  Although $P_q$ shows a
crossover transition and becomes vanishingly small at low $T$, the
free energy $F_q$ of a test charge $q$ remains finite at all $T$.
$P_q$ vanishes for a fractional charge.  However, one cannot conclude
that confinement of fractional charge from this result alone, as the
vanishing of $P_q$ follows solely from gauge invariance.

\begin{figure}[t,b]
\epsfxsize= 9.cm    
\epsffile[100 240 450 500]{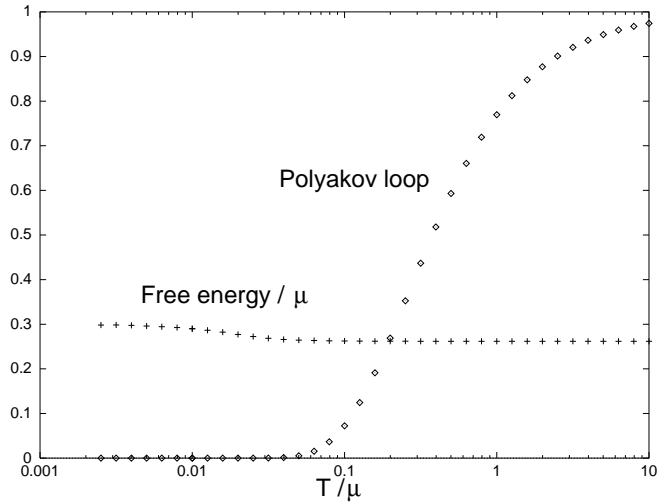}
\caption{$T$-dependence of the Polyakov loop and free energy in
the $N=3$ model with $m/\mu=.01$.  The free energy stays finite at all $T$.
\label{fig:Polyakov}}
\end{figure}

To get more information about confinement, one should insert a pair of
sources, one with charge $q$ and the other with $-q$, and examine the
increase (or decrease) of the energy.  The shift in the energy is
parametrised as $\Delta E = \sigma d + \cdots$ where $d$ is the
distance between the two sources, and $\sigma$ is the string tension.

In the multi-flavor case perturbation theory in mass cannot be employed.
Nevertheless one arrives at a simple result.  External charges are
completely screened, but the effective $\theta$ value is shifted between
the two sources by an amount $2\pi q/e$.  This shift in turn changes
the chiral condensate and therefore the energy density between the 
sources.\cite{CJS,HNZ,Gross,Grignani}

One finds
\beeq
\sigma = Nm \Big\{ \la \psibar\psi\ra_{\theta_\eff}
 - \la\psibar\psi\ra_\theta \Big\} ~~~,~~~
\theta_\eff = \theta-\myfrac{2\pi q}{e} ~. 
\label{string1}
\eneq
In particular, at $T=0$
\beeq
{\sigma  \over \mu^2}  = -  {N\over 2\pi} \, \Big( 2e^\gamma
\, {m\over \mu} \Big)^{{2N\over N+1}} 
\bigg\{ \Big( \cos {\bar\theta_\eff\over N}  \Big)^{{2N\over N+1}}
-\Big( \cos {\bar\theta\over N}  \Big)^{{2N\over N+1}} \bigg\}
\label{string2}
\eneq
Notice that the string tension vanishes for an integer $q/e$.
The string tension 
$\sigma$ is non-vanishing  only when fermions are massive
and chiral condensates have non-trivial $\theta$ dependence.

\section{Heavy fermions}
When fermion masses become large, we need to solve a more general
problem. For one flavor there is no $\vphi$ degree of freedom.
The vacuum wave function is expressed as $\hat f(p_W)$ which must satisfy
\beeq
\bigg\{  - {d^2\over d p_W^2}  + \omega^2 \, p_W^2
-  k \cos (2\pi p_W + \theta )  \bigg\} \hat f(p_W) = \ep \hat f(p_W) ~.
\label{Sch2}
\eneq
Here $\omega= \pi\mu L$ and $k$ depends on the fermion mass and the wave
function itself.  One finds that
\beeq
\la \psibar\psi\ra_{T=0} \sim 
\cases{ - \myfrac{\mu}{2\pi} \, e^\gamma
\cos\theta &for $m \ll \mu$\cr
 - \myfrac{m}{\pi}\, e^{2\gamma} &for $m \gg \mu$ .\cr}
\eneq

The chiral condensate becomes bigger as the fermion mass $m$
grows. The effect of a very heavy fermion never disappears even if $m$
approaches infinity. Does this contradict the decoupling theorem,
that heavy fermions are irrelevant in low energy physics?

The answer is `no'.  Chiral condensates of heavy fields are not good
measures of low energy physics.  Instead one should examine, for
instance,  the string tension defined when a pair of external
sources is inserted as was done in the previous section.

Eq.\ (\ref{string1}) shows that the string tension $\sigma$ depends on
two factors: $m$ and the $\theta$ dependence of $\la \psibar\psi\ra$.
Although $m\la \psibar\psi\ra$ becomes very large as $m$ increases,
its $\theta$ dependence diminishes rapidly for a large $m$.  In fig.\
3 we have displayed the $m$ dependence of $\sigma$ for $\theta=0$ and
$q/e=\onehalf$.  One can see that $\sigma$ increases for moderate $m$,
but quickly approaches zero as $m$ becomes large.  Extremely heavy
fermions are thus irrelevant for low energy physics.

\begin{figure}[t,b]
\epsfxsize= 9.cm    
\epsffile[100 240 450 500]{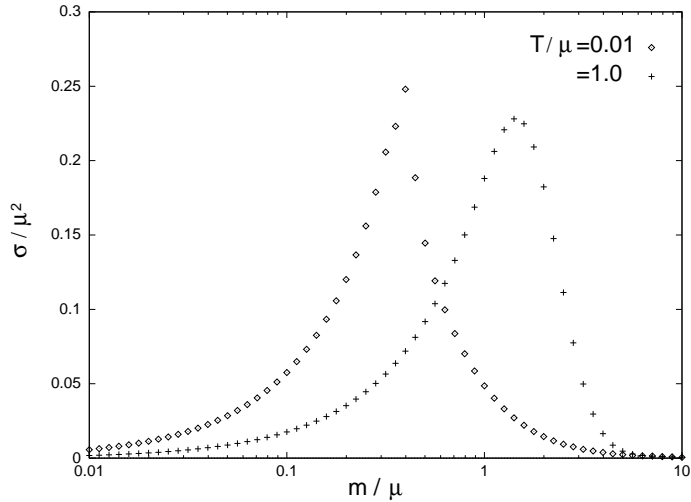}
\caption{Mass dependence of the string tension $\sigma$ in
the $N=3$ model.  $\sigma$ vanishes for a large fermion mass, which
is consistent with the decoupling theorem.
\label{fig:sigma}}
\end{figure}

\section{Summary}

There are similarities and differences between QCD$_4$ and QED$_2$:
both have confinement, and their chiral dynamics are pretty much the same.
However, in QED$_2$ a non-vanishing string tension (ie. confinement)
results only if $m\la \psibar\psi\ra$ has non-trivial $\theta$ 
dependence.   In other words, if there were no chiral condensates,
there would be no confinement in QED$_2$.

This can hardly be true in QCD$_4$ where it seems that confinement
and spontaneous chiral symmetry breaking are two separate phenomena.
Chiral symmetry is spontaneously broken even in the chiral limit $m_a=0$,
which is not the case in multiflavor QED$_2$. Nonetheless lattice
simulations show that confinement and chiral symmetry breaking are
intimately related.\cite{MILC}

Despite these differences, we can still learn quite a bit of QCD
physics from QED$_2$.  After all we have determined various physical
quantities such as chiral condensates, Polyakov loop, and string
tension at any temperature and with arbitrary fermion masses.  This is
the luxury of two-dimensional gauge theory. In the work summarized
here, we have presented a powerful method for solving QED$_2$ which
compliments lattice gauge theory and light-front
methods.\cite{lattice}

\section*{Acknowledgments}
This work was supported in part by the U.S.\ Department of Energy
under contracts  DE-AC02-83ER-40105
(Y.H.),   DE-FG02-87ER-40328 (R.R.), and  DE-FG03-95ER-40906 (J.H.)

\end{document}